\documentclass{article}

\usepackage[final]{nips_2017}

\usepackage[utf8]{inputenc} 
\usepackage[T1]{fontenc}    
\usepackage{hyperref}       
\usepackage{url}            
\usepackage{booktabs}       
\usepackage{amsfonts}       
\usepackage{nicefrac}       
\usepackage{microtype}      
\usepackage{graphicx,float}
\usepackage{amssymb,amsmath,amsthm}
\usepackage{cleveref}
\newcommand{\ignore}[1]{}

\title{Towards meaningful physics from generative models}

\author{
Marco Cristoforetti\thanks{Corresponding author} $\quad$ Giuseppe Jurman $\quad$ Andrea I. Nardelli $\quad$ Cesare Furlanello\\ 
Fondazione Bruno Kessler\\
Trento, Italy\\
\texttt{ \{mcristofo,jurman,anardelli,furlan\}@fbk.eu}
}

\begin{document}
\maketitle

\begin{abstract}
In several physical systems, important properties characterizing the system itself are theoretically related with specific degrees of freedom.
Although standard Monte Carlo simulations provide an effective tool to accurately reconstruct the physical configurations of the system, they are unable to isolate the different contributions corresponding to  different degrees of freedom.
Here we show that unsupervised deep learning can become a valid support to MC simulation, coupling useful insights in the phases detection task with good reconstruction performance.
As a testbed we consider the 2D XY model, showing that a deep neural network based on variational autoencoders can detect the continuous Kosterlitz-Thouless (KT) transitions, and that, if endowed with the appropriate constrains, they generate configurations with meaningful physical content.
\end{abstract}

\section{Introduction}
\label{sec:intro}
In physics, actual understanding of phenomena often heavily relies on the feasibility of the computational analysis, \textit{i.e.} the possibility of performing effective numerical simulations of the theory. 
These simulations generate different \textit{microstates} of a theoretical model of a real thermodynamic system, from which the macroscopic observable quantities are computed.
For a thermodynamic system in a given thermal state, a microstate is a microscopic (\textit{i.e.}, describing the state of each element of the system) configuration that the system may occupy with a certain probability.
On the other hand, a \textit{macrostate} is defined by one or more overall macroscopic properties, such as temperature, pressure, energy, etc.
There are thus several possible microstates accounting for the same macrosystem.
As a rule of thumb, a key aspect is the characterization of dramatic changes occurring among set of configurations representing different system macrostates: these changes are known as \textit{phase transitions}.
The standard approach uses Monte Carlo methods to sample configurations from the \textit{partition function} (the function that represents the theory of the model) and on these configurations expectation values of observables are calculated. 
However, this computational strategy is prone to several drawbacks, the two most critical being:
\begin{enumerate}
\item To detect phase transitions, an order parameter, a quantity whose dramatic variation as a function of the temperature signals the transformation, must be defined. 
In many systems, defining this kind of observable can be a hard task. 
\item System properties can be associated to specific degrees of freedom, \textit{e.g.}, topological excitations. 
It is hard or even impossible with current techniques to isolate such contributions from Monte Carlo generated configurations.
\end{enumerate}

Statistical and computational physics has majored as an interesting playground for the machine learning community, 
(see for instance \cite{srinivasan13revisiting} \cite{rezende14stochastic, torlai16learning, schindler17probing, carrasquilla17machine}), 
achieving relevant results and with reciprocal benefits, as theoretical insights in learning that can be exploited by using physical approaches (\cite{mehta14exact}).
In particular, recent works in the literature tackle the problem of discovering and interpreting phase transitions via machine learning, \textit{e.g.} \cite{hu17discovering,vannieuwenburg17learning,wang16discovering,tanaka17detection}, while other authors focus on substituting Monte Carlo simulations with an equivalent learning approach (see \cite{huang17accelerate,liu16self}): in both cases, the arise of the deep learning paradigm is strongly contributing to speed up this trend.

This work follows such line of thought, aiming to apply deep learning algorithms on top of Monte Carlo simulations. 
In this context the 2D XY model, detailed in Sec.~\ref{sec:xy}, has all the properties for exploring the power of machine learning for phase transition identification and investigation of the internal structure of the Monte Carlo configurations.

Characteristic of the 2D XY model is a continuous transition, not related with any symmetry breaking, called Kosterlitz-Thouless (KT) transition (\cite{kosterlitz73ordering}). 
The use of deep learning methods to investigate this type of transformation extends previous research works where machine learning has been applied to identify true phase transitions, by using both supervised as well as unsupervised techniques  (\cite{aoki16restricted,portman16sampling,wetzel17unsupervised}).

The KT transition in this model is mediated by specific topological defects (vortices) that unbind at high temperature; therefore this is a perfect situation to prove the effectiveness of deep learning also in order to identify the relevance of specific degrees of freedom. 

Operatively, we adopt the Variational AutoEncoders (VAE) network for our study (\cite{doersch16tutorial,kingma14autoencoding}), obtaining promising results focusing on the ability of generative models to reconstruct configurations with meaningful physics.

\section{The XY model}
\label{sec:xy}
In the 2D XY model, the degrees of freedom are planar rotors ($\vec{\sigma}$) of unit length arranged on a two dimensional square lattice.
Here the rotors take continuous values in $[0,2\pi)$, differently from the classical Ising model where rotors are seen as binary spins assuming only two positions up or down, and for which analysis is simpler (see \cite{portman16sampling}).

For the 2D XY model, the Hamiltonian of the system is given by
\begin{equation}\label{eq:hamiltonian}
H_{XY} = -J \sum_{\langle i,j\rangle}\vec{\sigma}_i\cdot\vec{\sigma}_j = -J \sum_{\langle i,j\rangle}\cos(\theta_i-\theta_j)\ ,
\end{equation}
where $\langle i,j\rangle$ denotes all adjacent sites on the lattice and $\theta_i$ denotes the angle of the rotor on site $i$.

The Mermin-Wagner-Hohenberg theorem (\cite{mermin66absence}) prohibits continuous phase transitions in $d \leq 2$ dimensions at finite temperature when all interactions are sufficiently short-ranged. 
Nevertheless the 2D XY model shows a Kosterlitz-Thouless transition (\cite{kosterlitz73ordering}) connected with the presence of topological charges in the system. 
In particular, vortices that are bounded in the ordered phase, below the critical temperature $T_c$, unbind at high temperatures causing an unordered phase where the correlation function decays exponentially.
This kind of transition is therefore not related with any broken symmetry and it is meaningful to see if it can be captured by Deep Learning networks.

In the classical approach, configurations are generated by using Markov chain Monte Carlo to sample the probability distribution:
\begin{displaymath}
P_\theta(T)=\frac{\exp^{-\beta H(\theta)}}{Z(T)}\ ,
\end{displaymath}
where $\theta$ represents a given configuration of the system, $\beta = 1/k_B T$ for $k_B$ the Boltzmann constant, $T$ is the temperature and $Z$ is the partition function at a given temperature defined by
\begin{displaymath}
Z(T) = \sum_\theta \exp^{-\beta H(\theta)}.
\end{displaymath} 
Hereafter we consider a system with linear size $L=16$ which allows the study of all the characteristics of the model despite the small dimension.
Operatively, $10.000$ uncorrelated configurations are sampled, for $\beta$ linearly increasing from $0.1$ to $1.9$ by $0.1$ steps.

\section{VAE for the 2D XY model}
\label{sec:vae}
Variational autoencoders (VAE) are generative models, learning the parameters of a probability distribution by modeling the data through autoencoders (\cite{kingma14autoencoding}). 
After learning the probability distribution, parameters can be sampled so that the encoder network can generate samples closely resembling the training data. 
In particular, here we consider the Convolutional VAE that has been shown to be more effective in the encoding of systems of the class considered in this paper (\cite{wetzel17unsupervised}).
The VAE has four convolutional layers, two fully connected layers and the layer of the latent variables, which will be later shown to be sufficient to correctly capture the phase transition.

\begin{figure}[!b]
\begin{center}
\begin{tabular}{cc}
(a) & \includegraphics[width=0.8\textwidth]{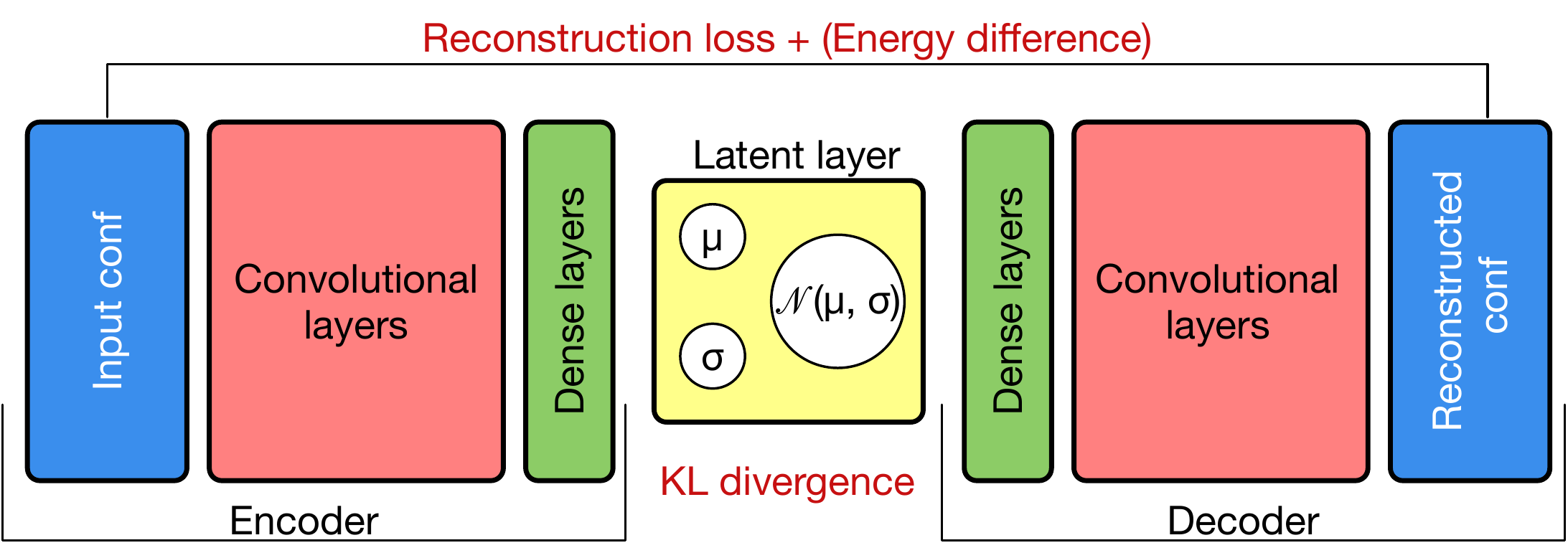} \\
(b) & \includegraphics[width=0.8\textwidth]{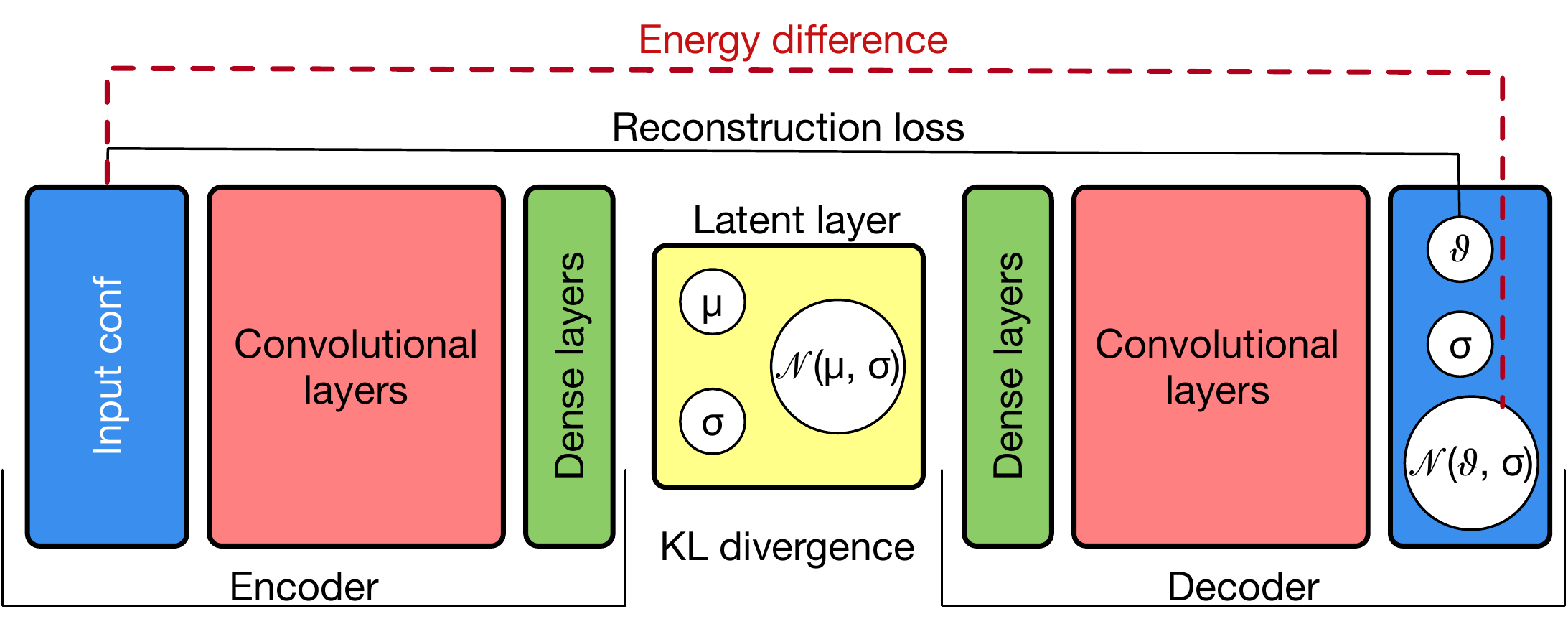} 
\end{tabular}
\end{center}
\caption{Architectures of the HG-VAE: (a) including the energy difference between the original and reconstructed configuration; (b) splitting the last layer in two terms representing the parameters of the Gaussian distribution used to generate the configurations for evaluating the energy loss term.}
\label{fig:vaes}
\end{figure}

Adopting an incremental approach from a basic architecture, the standard VAE is first tested using the sum of reconstruction loss and Kullback-Leibler loss as objective function.
As our goal goes beyond the identification of the phase transition towards the development of a method to investigate collective dynamics in the configurations. 
To this aim, we first show that the Standard VAE is not sufficient, thus we enrich the architecture by introducing two novel ingredients, as graphically detailed in Fig.~\ref{fig:vaes}, that will be later shown to considerably improve the physical meaning of the reconstructed configurations.
From now on we indicate this new network as HG-VAE.
First, some physical insight is injected in the VAE by modifying the loss function and adding a new term measuring the energy difference between the original and the reconstructed configurations:
\begin{displaymath}
\textrm{loss}_H = \left(E\left(\theta_{MC}\right) - E\left(\theta_R\right)\right)^2
\end{displaymath}
where $\theta_{MC}$ and $\theta_{R}$ are the Monte Carlo and reconstructed configurations, respectively (Fig.~\ref{fig:vaes}, upper panel).

Further, the reconstruction layer is split into two terms $\theta$ and $\sigma$ corresponding to the parameters of a Gaussian distribution. 
In detail, by this position there are two quantities associated with each element of the configuration: $\theta_i$, the output of the standard VAE, used to evaluate the reconstruction loss and a standard deviation $\sigma_{\theta_i}$. 
A configuration extracted from the sampling of the Gaussian $\mathcal{N}(\theta_i,\sigma_{\theta_i})$ is then considered for the evaluation of the $\textrm{loss}_H$ term (Fig.~\ref{fig:vaes}, lower panel). 

\section{Results}
\label{sec:results}

\begin{figure}[!t]
\centering
\begin{minipage}[c]{.50\textwidth}
\centering
\includegraphics[width=1\textwidth]{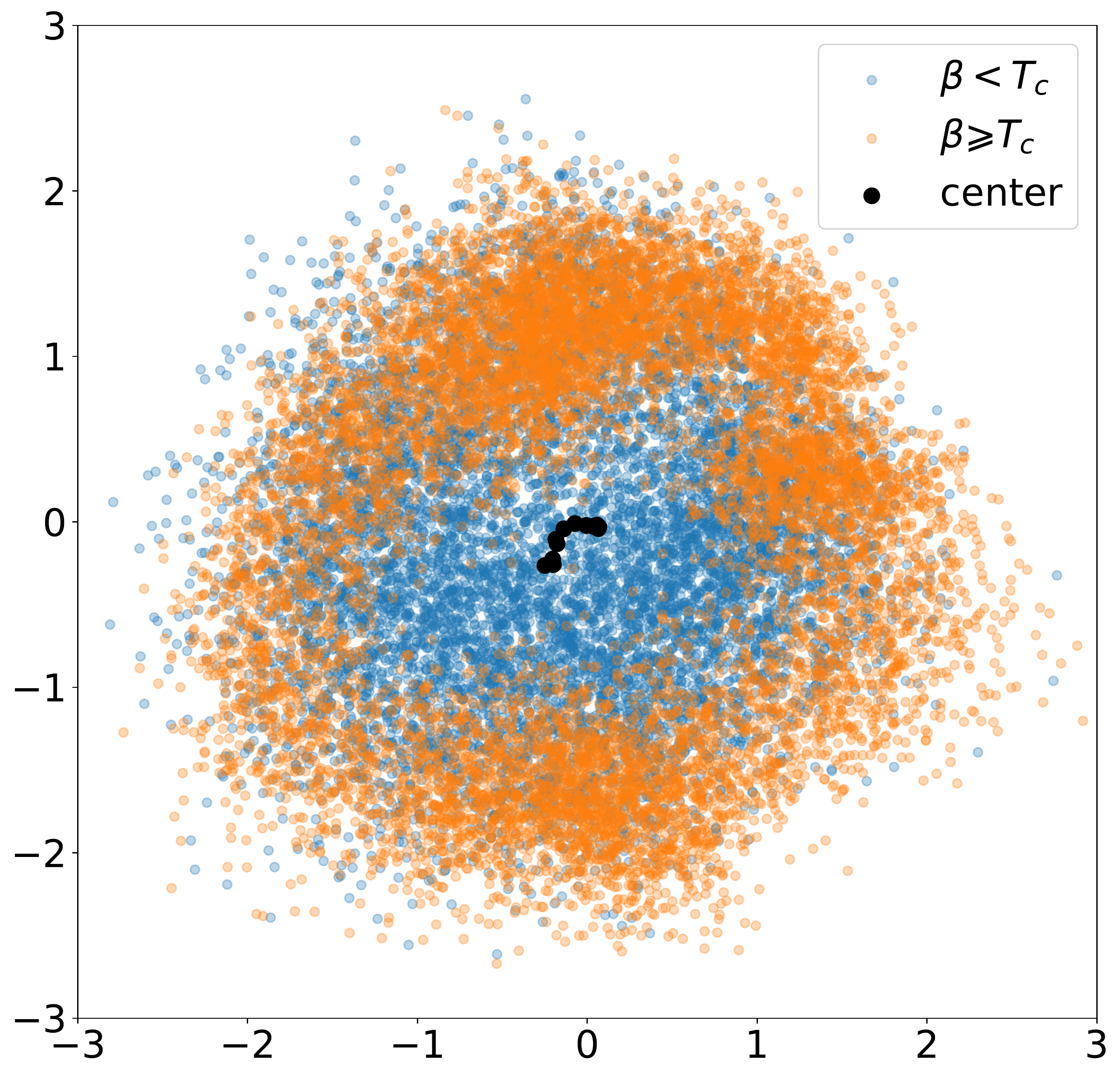}
\end{minipage}%
\hspace{0mm}%
\begin{minipage}[c]{.50\textwidth}
\centering
\includegraphics[width=1\textwidth]{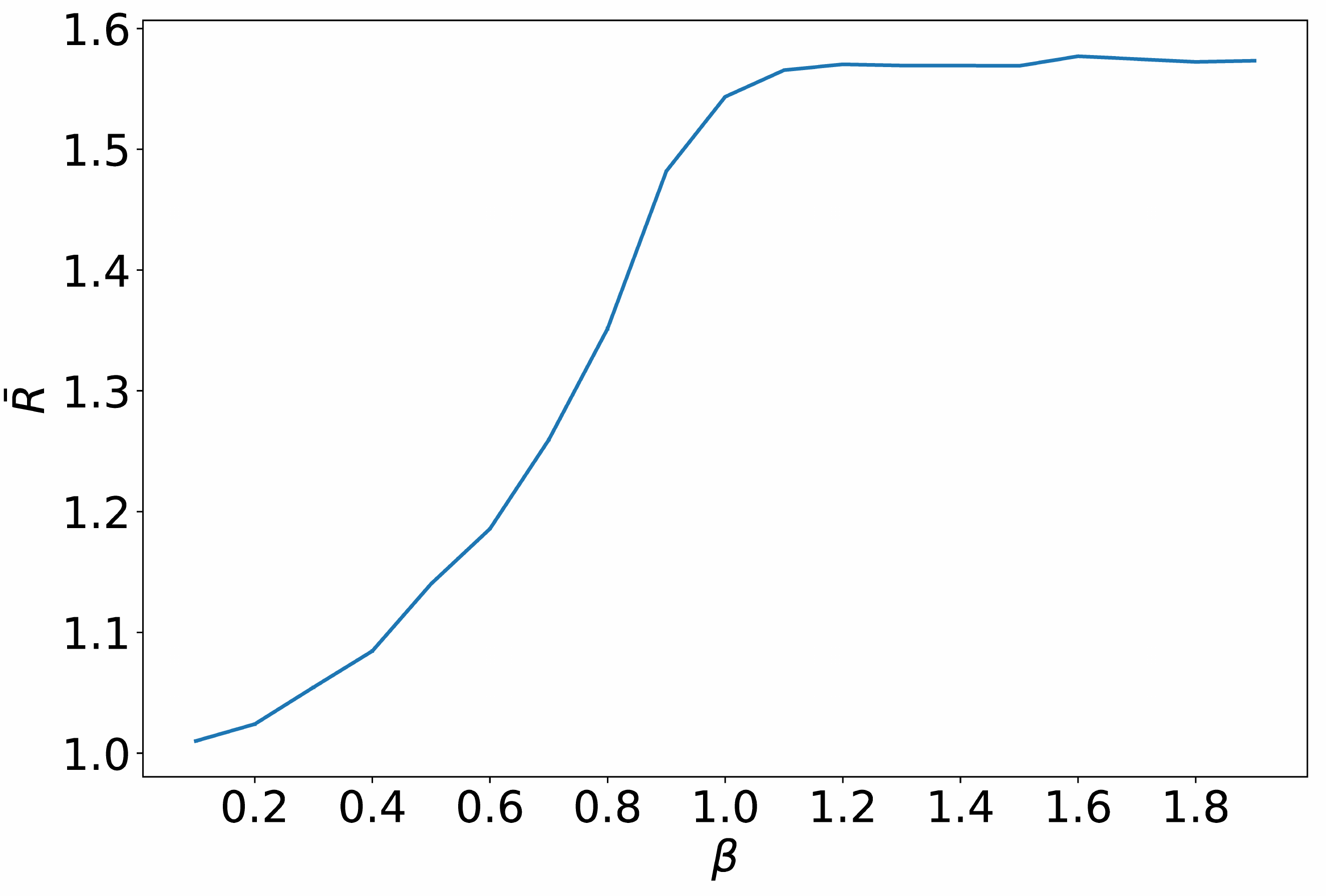}\\
\includegraphics[width=1\textwidth]{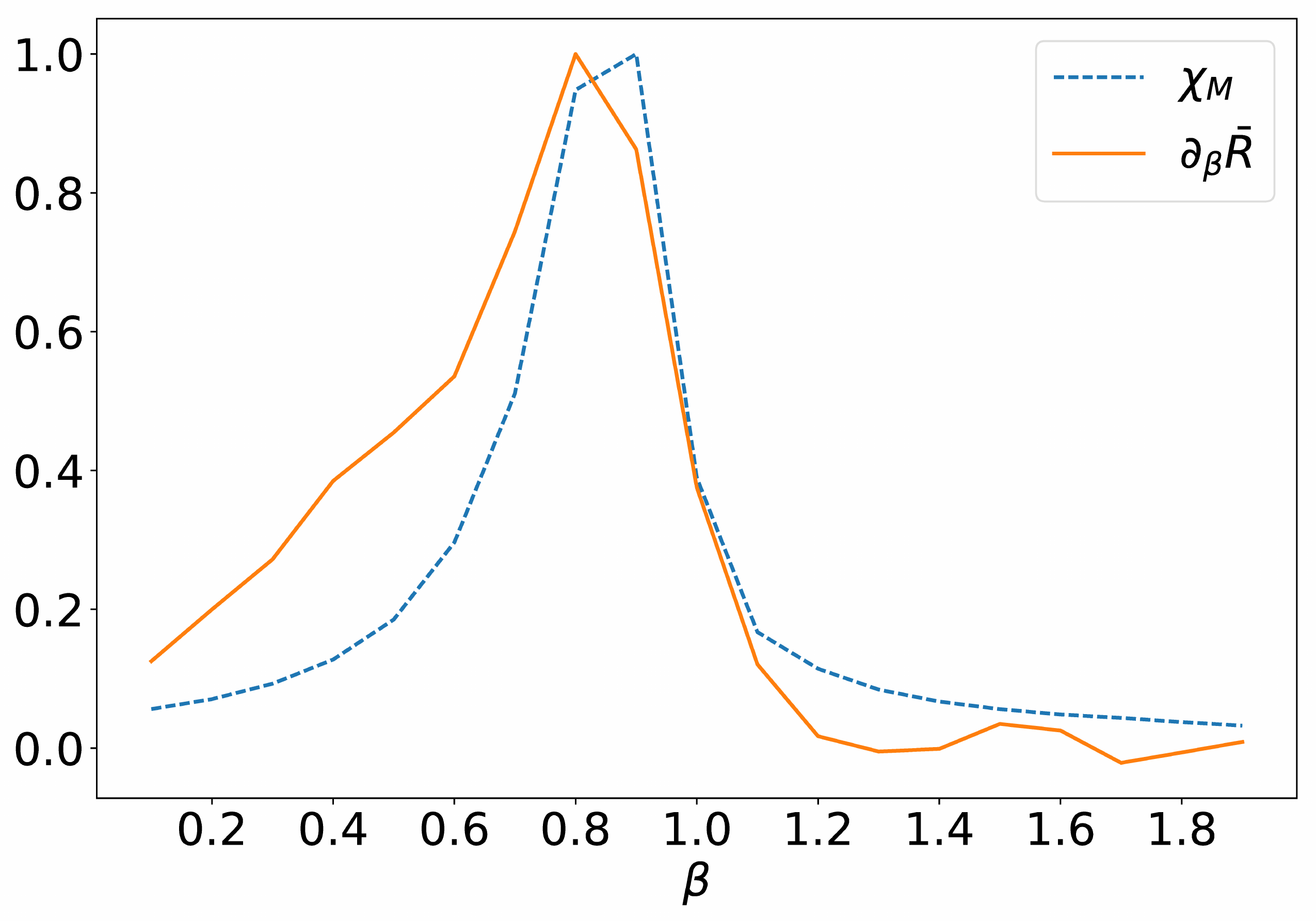}
\end{minipage}
\caption{Detection of the phase transition using the Standard VAE for the 2D XY model: (a) Distribution of the latent variables obtained from the Monte Carlo generated configurations at all $\beta$s. (b) $\bar{R}$ as order parameter of the transition. (c) Comparison of the magnetic susceptibility and the derivative of $\bar{R}$. The critical point detected from the VAE is quite close to the true one.}
\label{fig:lv_vae}
\end{figure}

\subsection{Standard VAE}
\label{ssec:standard}
The Standard VAE is trained with configurations generated by using Markov chain Monte Carlo, sampling every 100 steps in order to avoid autocorrelations, for $19$ values of $\beta$ in the range $[0.1,1.9]$ with a $\beta$ step of $0.1$.
For models with phase transitions connected with a symmetry breaking, there is evidence that the transition can be clearly identified by looking at the distribution of the latent variables  (\cite{vannieuwenburg17learning}).
As anticipated before, the XY model is different because the transition is connected with the unbinding of vortices, and a priori it is unclear whether the VAE can capture this kind of transformation.
In Fig.~\ref{fig:lv_vae} we show that even in this case latent variables corresponding to configurations with $\beta<\beta_c$ distribute differently from those at $\beta \geq \beta_c$, with $\beta_c$ the critical value obtained from the Monte Carlo simulations.
Encoding 1000 configurations for each $\beta$ we have a distribution of the latent variables plotted in Fig.~\ref{fig:lv_vae}-(a). 
The two colors refer to values of $\beta$ smaller (blue) or larger (orange) than the critical point. 
For large $\beta$s, corresponding to lower temperatures and thus to the ordered phase, the latent variables distribute on a subset of the space occupied from latent variables at small $\beta$s and far from the center of the distribution.
In the same plot, the black dots point the center of the distribution at each $\beta$. 
If the mean Euclidean distance $\bar{R}$ of the latent variable from the related center of the distribution is computed for all configuration for each $\beta$, this quantity can be treated as an order parameter for the transition (Fig.~\ref{fig:lv_vae}-(b)).
Comparing the derivative of $\bar{R}$ with the magnetic susceptibility obtained from the Monte Carlo configurations, and defined by
\begin{displaymath}
	\chi = V\left(\langle m^2\rangle - \langle m\rangle^2\right),
\end{displaymath}
we see that the critical point identified from the VAE analysis approximate the true point with good accuracy (Fig.~\ref{fig:lv_vae}-(c)).
From this analysis we can conclude that the Standard VAE, regardless of any physical insight, can successfully be applied to the identification of phase transitions,  not only when these transformations are connected with a symmetry breaking but also when these are continuous such as the KT transition.

As far as the ability of unsupervised algorithms to generate physically meaningful configuration is concerned, the Standard VAE should be able to reproduce extensive quantities like the energy, directly given by the Hamiltonian Eq.~\ref{eq:hamiltonian}, and the total magnetization:
\begin{displaymath}
	m(\vec{\sigma}) = \frac{1}{V}\sum_\mathbf{x}\vec{\sigma}_\mathbf{x}
\end{displaymath}

\begin{figure}[t]
\begin{center}
\begin{tabular}{cc}
\includegraphics[width=0.50\textwidth]{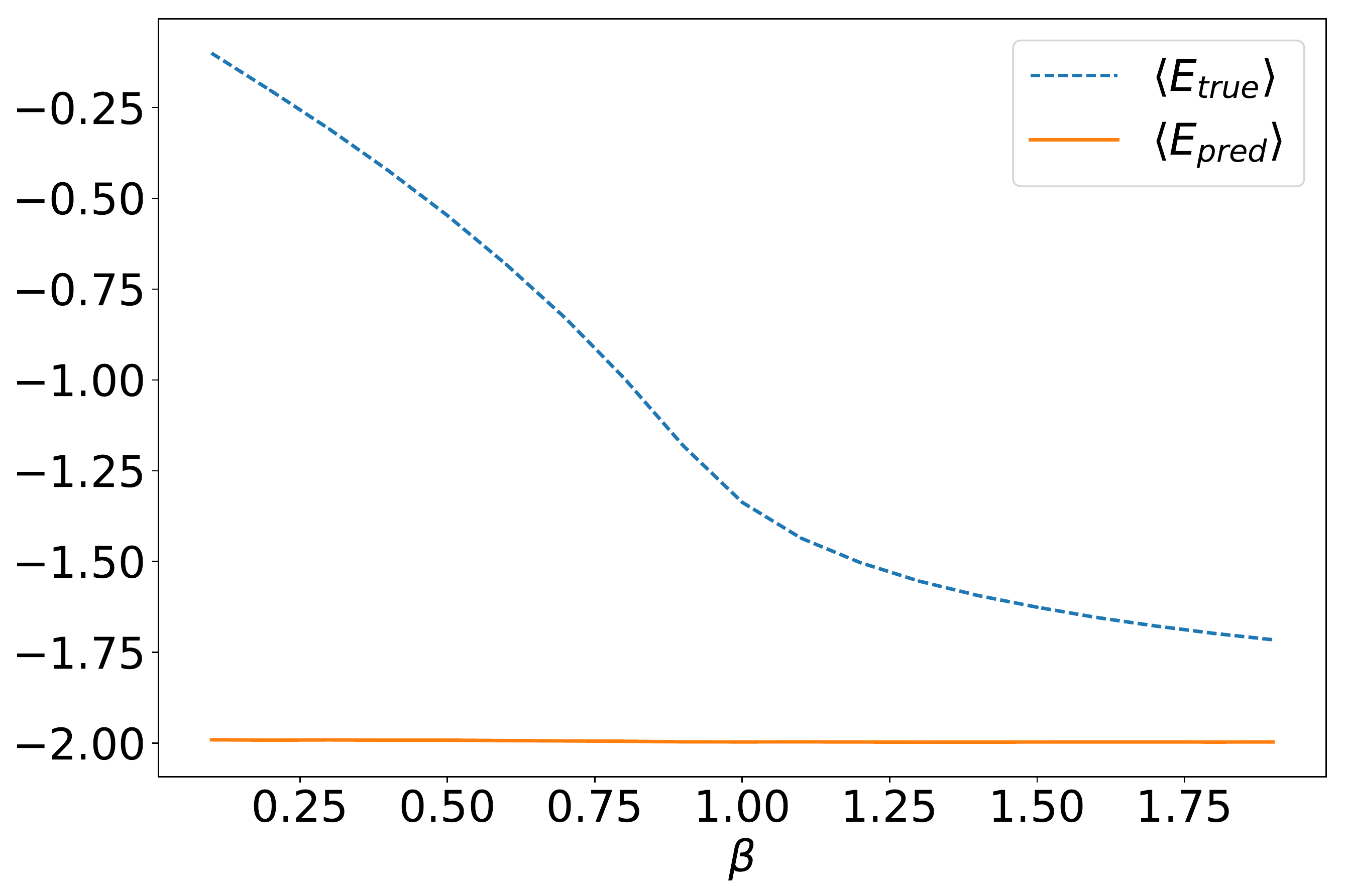} &
\includegraphics[width=0.48\textwidth]{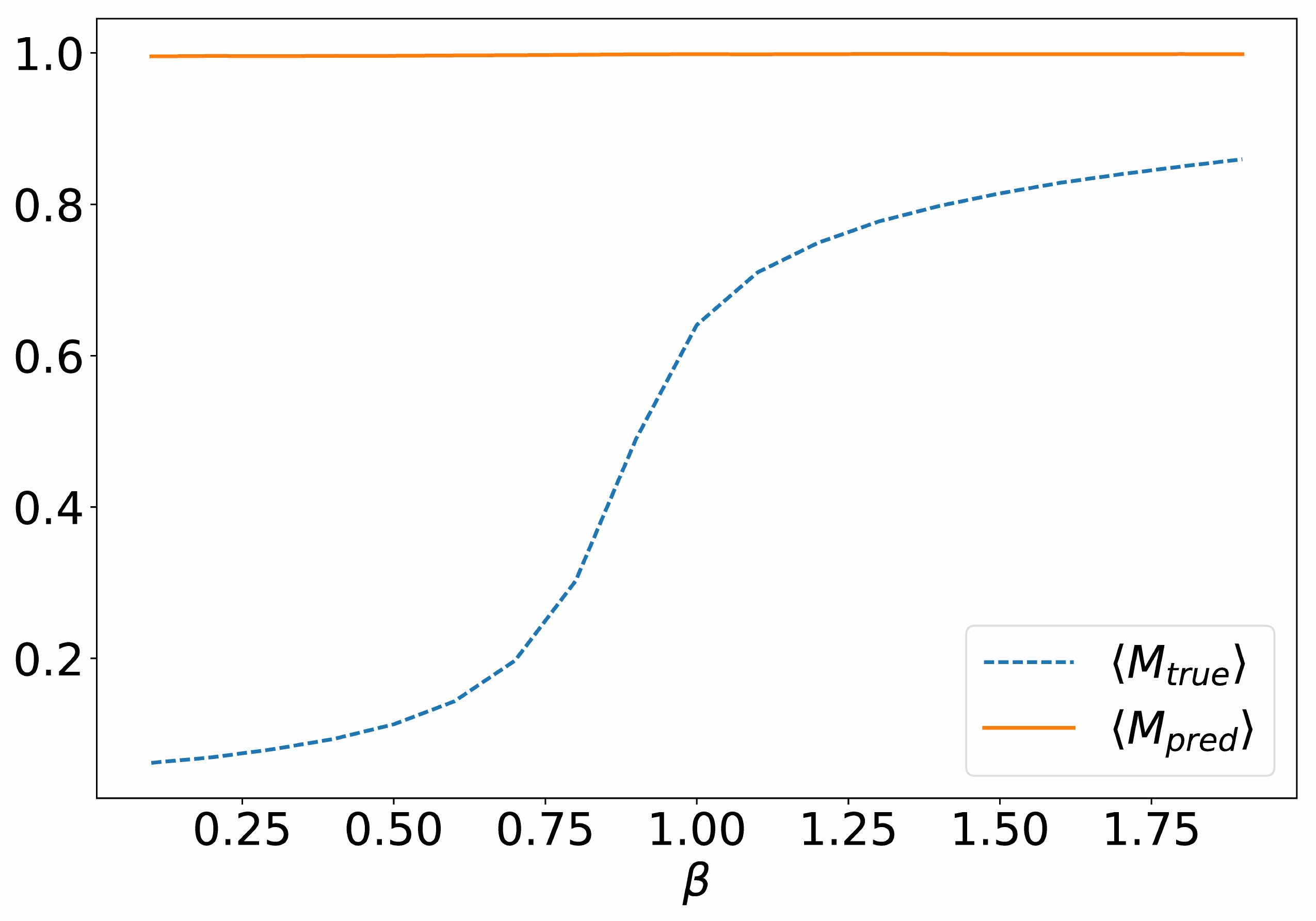} \\
(a) & (b)
\end{tabular}
\end{center}
\caption{Comparison of expectation values of (a) energy and (b) magnetization computed on the Monte Carlo (blue) and VAE reconstructed (orange) configurations.}
\label{fig:mag_en_vae}
\end{figure}

\begin{figure}[!ht]
\begin{center}
\includegraphics[width=0.95\textwidth]{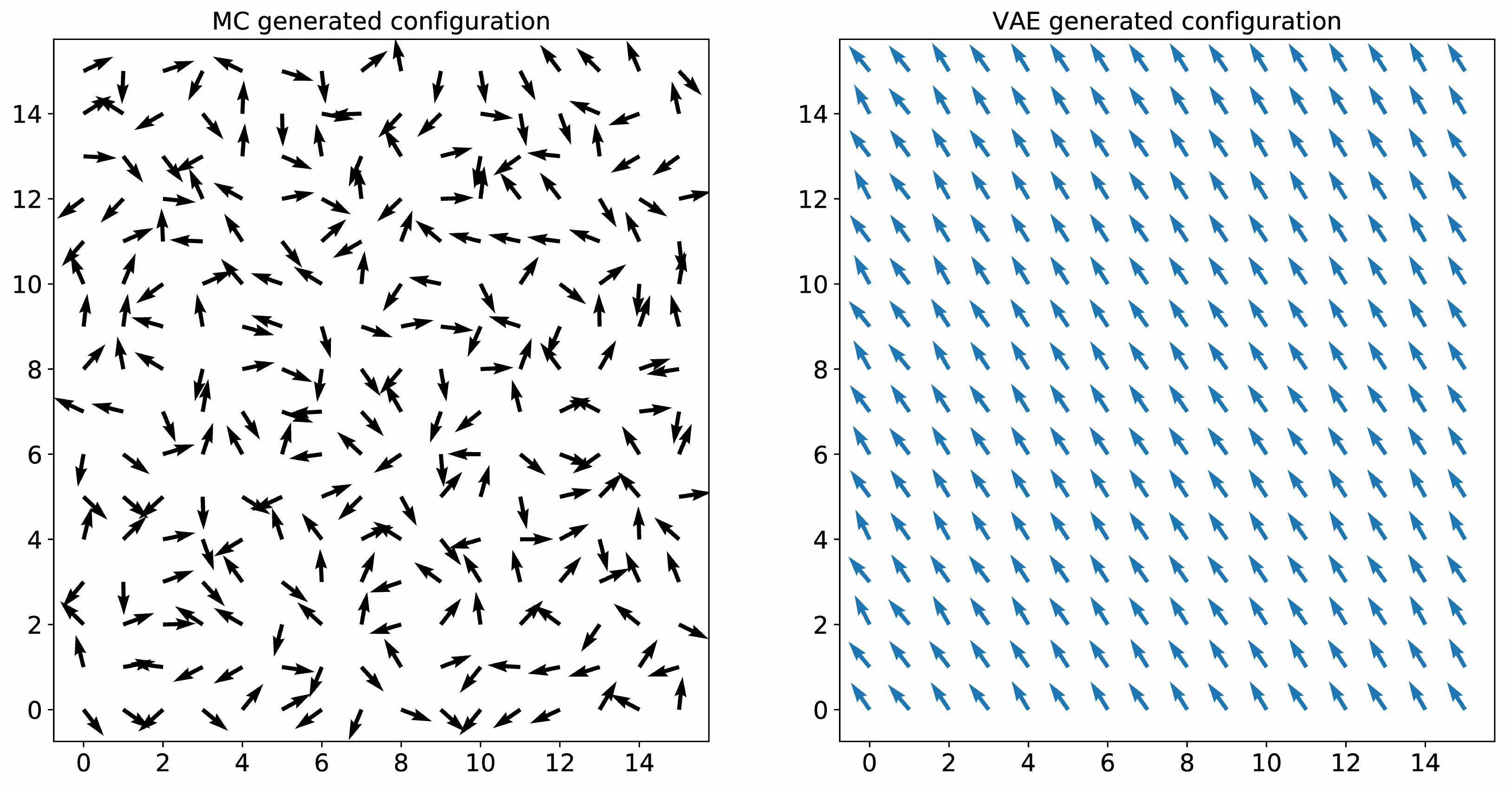}
\end{center}
\caption{Example of configuration reconstruction for $\beta=0.1$. The original configuration on the right, the reconstructed on the left.}
\label{fig:conf_ex_01}
\end{figure}

The expectation values of the energy and the magnetization, as computed on the Monte Carlo and the reconstructed configurations, are presented in Fig.~\ref{fig:mag_en_vae}.
It is clear that the Standard VAE fails completely to reproduce the correct value of the observables predicting a constant value as a function of $\beta$ which corresponds to the maximally ordered state of the system.
A visual inspection of a reconstructed configuration can help understanding the problem: in Fig.~\ref{fig:conf_ex_01} an example is displayed for $\beta=0.1$, \textit{i.e.}, the maximally disordered case (left: original, right: VAE configuration). 
The spins in the reconstructed case are all aligned on the same direction, meaning that the Standard VAE could not reproduce any specificity of the input configuration. 

\subsection{H/G-VAE}
\begin{figure}[t]
\begin{center}
\begin{tabular}{cc}
\includegraphics[width=0.5\textwidth]{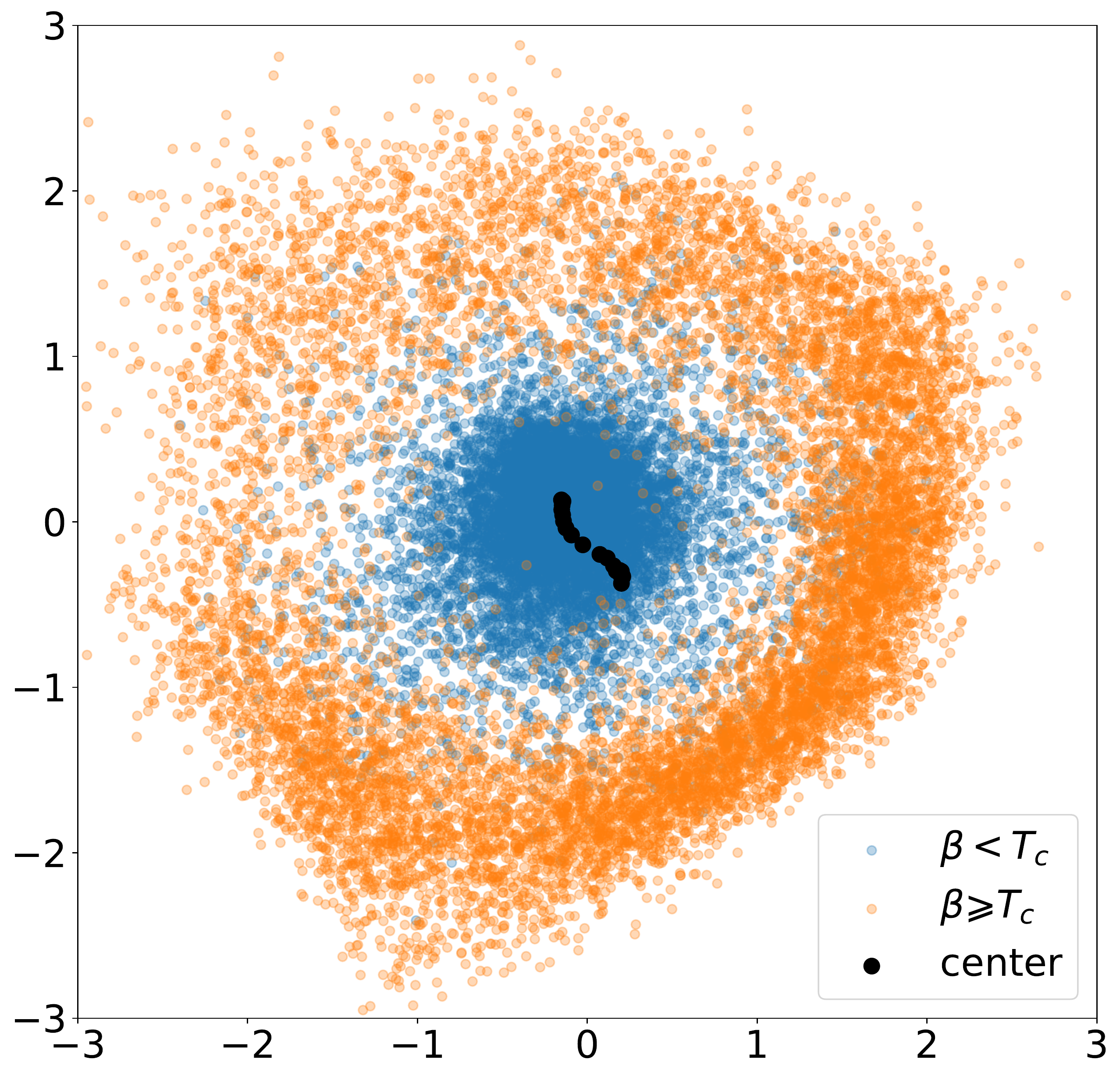} &
\includegraphics[width=0.5\textwidth]{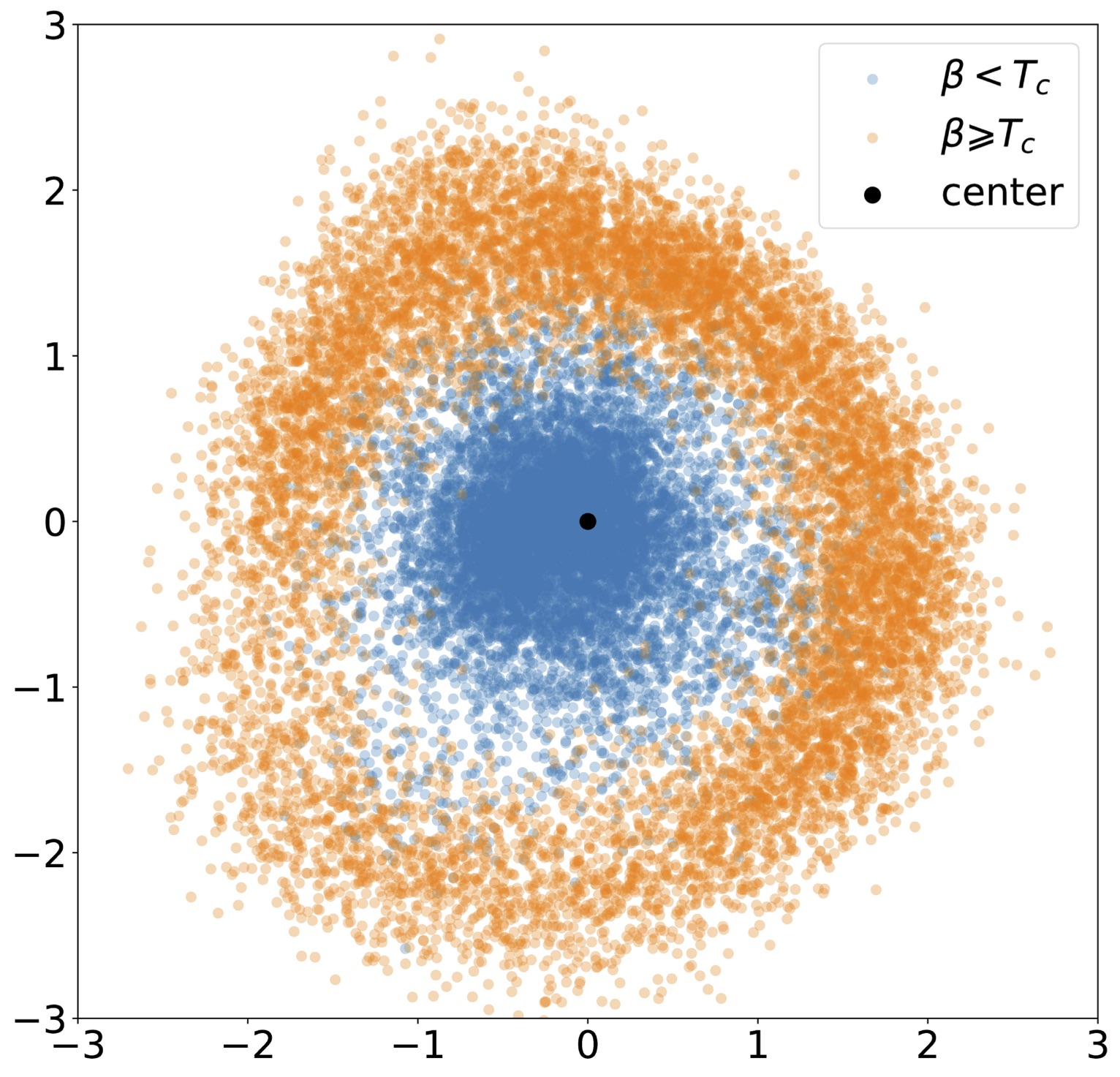} \\
(a) reconstruction + energy & (b) gaussian fluctuations
\end{tabular}
\end{center}
\caption{Distribution of the latent variables obtained using the HG-VAE. In both cases (a) with the energy loss term, (b) including Gaussian fluctuations, an improvement is achieved in the identification of the phase transition.}
\label{fig:lat_var_ext}
\end{figure}

\begin{figure}[b]
\begin{center}
\begin{tabular}{cc}
\includegraphics[width=0.5\textwidth]{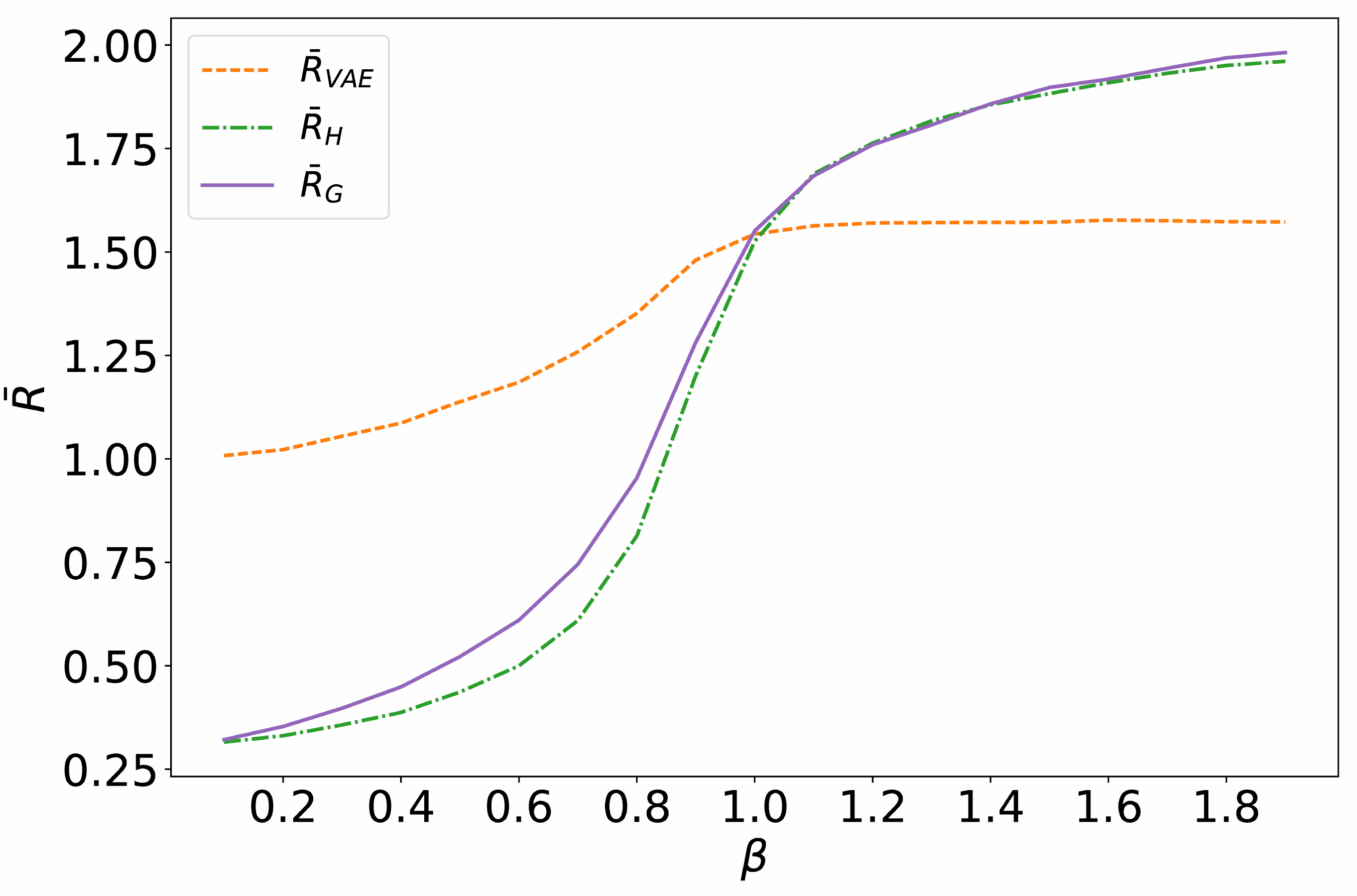} &
\includegraphics[width=0.48\textwidth]{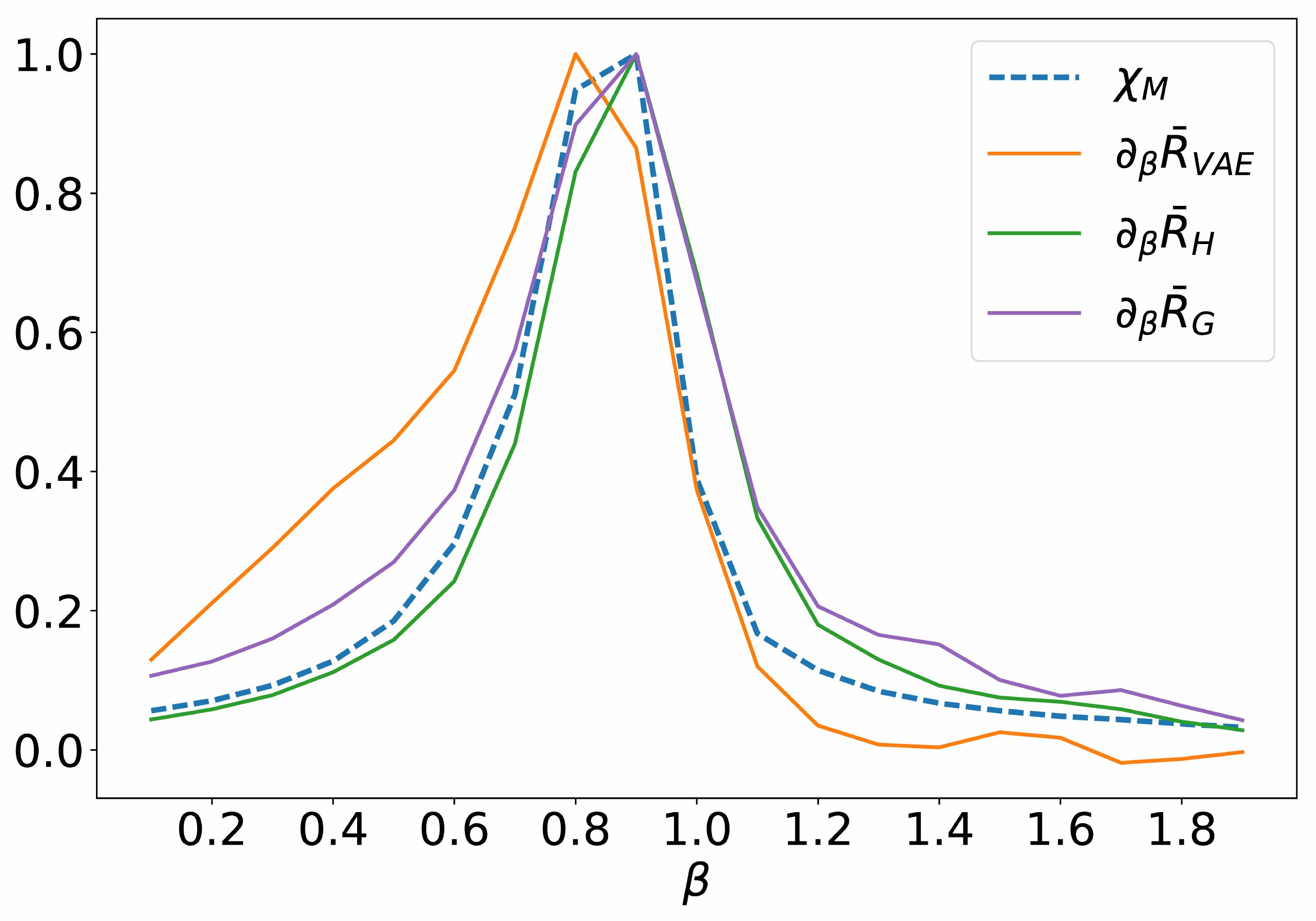} \\
(a) & (b) 
\end{tabular}
\end{center}
\caption{(a) Order parameter $\bar{R}$ in the three models (b) Comparison of the derivative of $\bar{R}$ obtained from the three VAEs and the magnetic susceptibility computed on the Monte Carlo configurations.}
\label{fig:order_p_comp}
\end{figure}

Trying to improve the reconstruction task we add physically driven modifications to the learning algorithm. 
This certainly breaks the view of Deep Learning as an agnostic approach which can be applied to different systems without contextual knowledge; as our focus is on the specific physical problem of detecting and isolating specific degrees of freedom inside the Monte Carlo generated configurations, this design is absolutely plausible.
Here we present the results obtained using the two physically motivated versions of the VAE introduced in Sec.~\ref{sec:vae}. 
In all the following figures, VAE indicates the Standard variational autoencoder, $H$ refers to the VAE with the energy loss term (H-VAE) and $G$ to the one including Gaussian fluctuations (G-VAE).
As a sanity check, we first check that the introduced physical constraint does not alter the ability of the VAE to identify the phase transition. 
In Figs.~\ref{fig:lat_var_ext} and ~\ref{fig:order_p_comp} we reproduce the analysis of the latent variables as in Sec.~\ref{ssec:standard} showing that, as expected, the H-VAE increases the ability of detecting the transformations. 
The distribution of the latent variables in the two cases $\beta<\beta_c$ and $\beta>=\beta_c$, clearly shows a separation in two regions corresponding to the two different phases. 
Moreover, the critical point now coincides with the true one as identified from the magnetic susceptibility.

\begin{figure}[!t]
\begin{center}
\begin{tabular}{cc}
\includegraphics[width=0.5\textwidth]{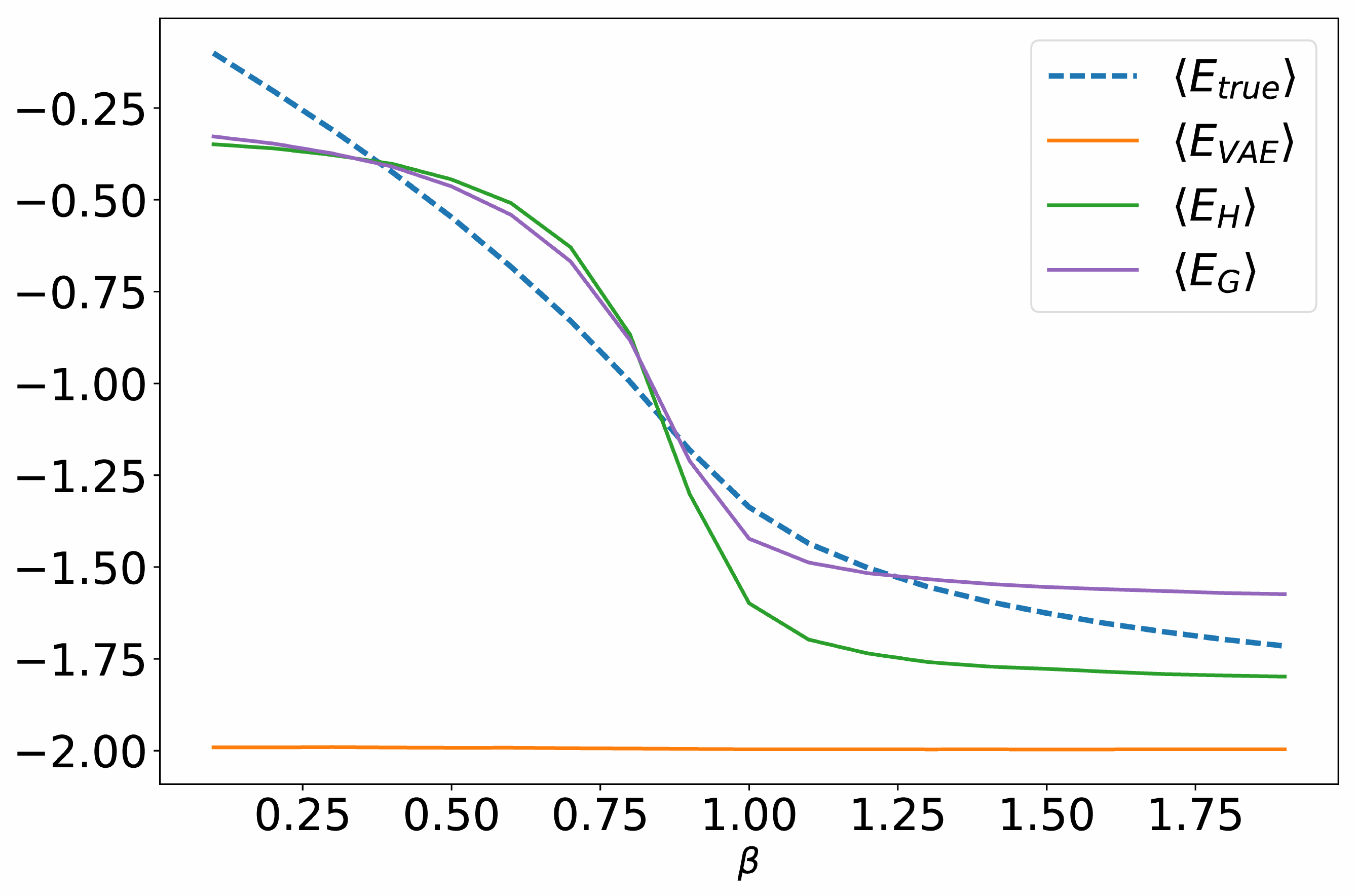} &
\includegraphics[width=0.47\textwidth]{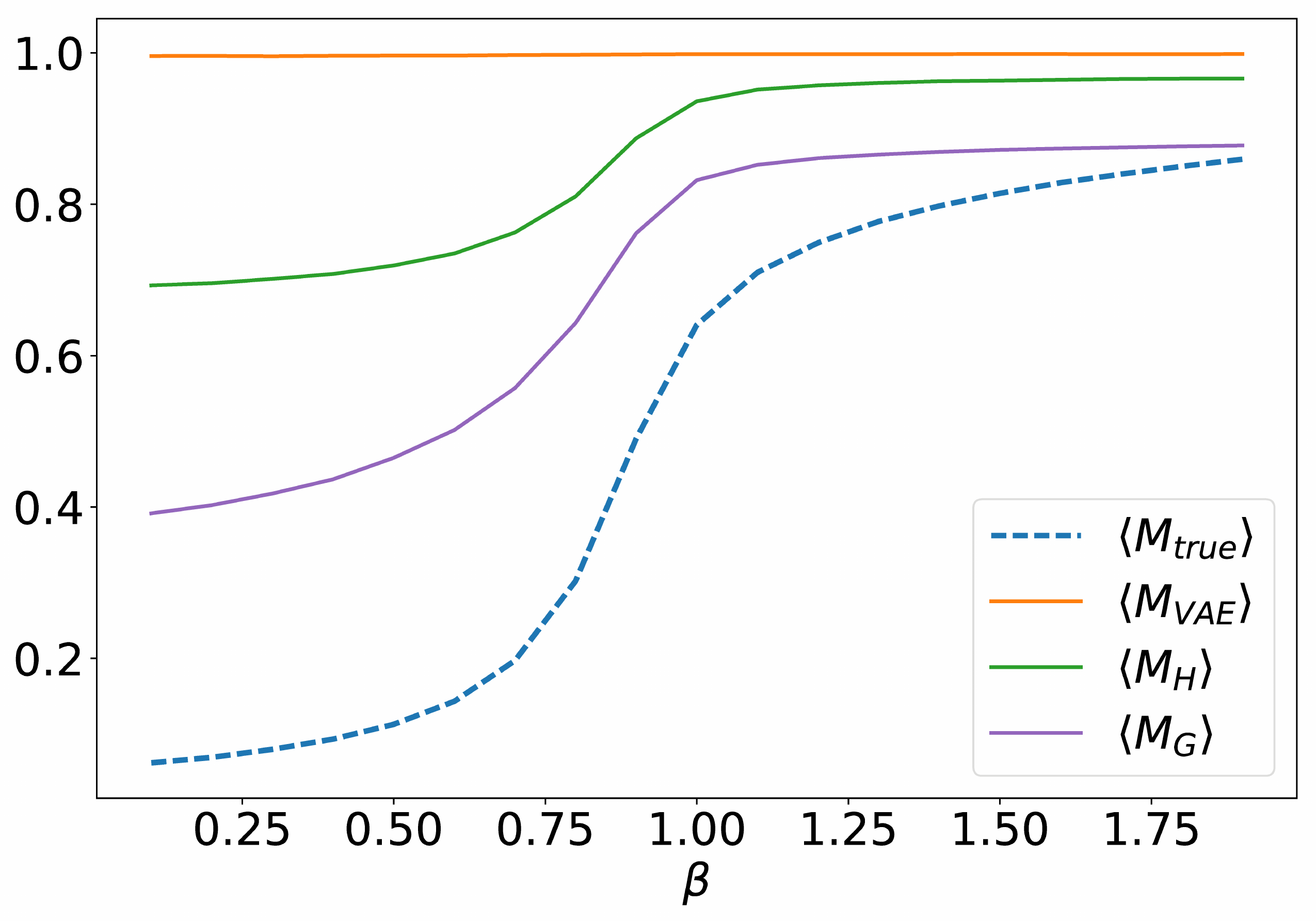} \\
(a) & (b) 
\end{tabular}
\end{center}
\caption{Energy (a) and magnetization (b) for the three version of the VAE compared to the true values obtained by Monte Carlo.}
\label{fig:energy_magn_comp}
\end{figure}

Including the proposed H and G terms in the Standard VAE improves the reconstruction performance reflected in the evaluations of the energy and magnetization (Fig.~\ref{fig:energy_magn_comp}).
There are still discrepancies between the true values and the generated ones. In future works we will try to amend this specialising VAEs to reproduce a single $\beta$ value. This strategy can probably improve the performance of the models.

\begin{figure}[htbp]
\begin{center}
\includegraphics[width=0.5\textwidth]{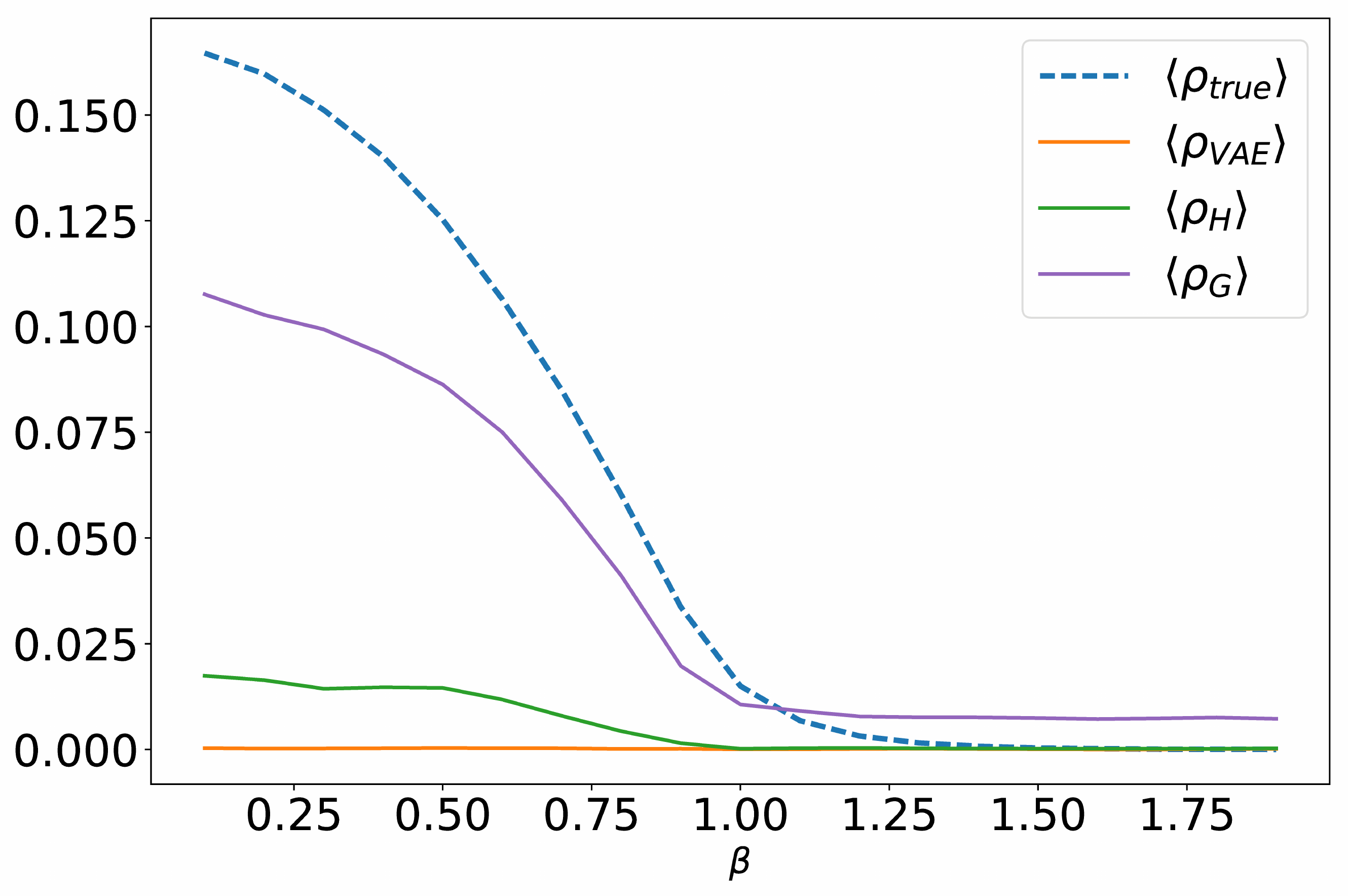}
\end{center}
\caption{Vortex density comparison between Standard VAE, H/G-VAE and true values.}
\label{fig:vort_density}
\end{figure}

As a further check, we compute the vortex density to verify the crucial effect of the G and H terms for the reconstruction task. Vortex density is defined by:
\begin{eqnarray*}
	&&\rho = \langle l_ij\rangle = \frac{1}{V} \sum_\mathbf{x}\sum_{i\neq j}|l_{ij,\mathbf{x}}|\\
	&&l_{ij,\mathbf{x}} = \frac{1}{2\pi}(n_{i,\mathbf{x}} + n_{j,\mathbf{x}+\mathbf{i}} - n_{i,\mathbf{x}+\mathbf{j}} - n_{j,\mathbf{x}})\\
	&&n_{i,\mathbf{x}} = (\theta_{\mathbf{x}+\mathbf{i}}-\theta_\mathbf{i}).
\end{eqnarray*}
From Fig.~\ref{fig:vort_density} we see that by including the Gaussian fluctuations, the vortex density is much larger and moves significantly towards the true value. 
For large $\beta$, the Gaussian contribution is probably too large, allowing the formation of vortices in a region where the density should be zero.

\begin{figure}[htbp]
\begin{center}
\includegraphics[width=1\textwidth]{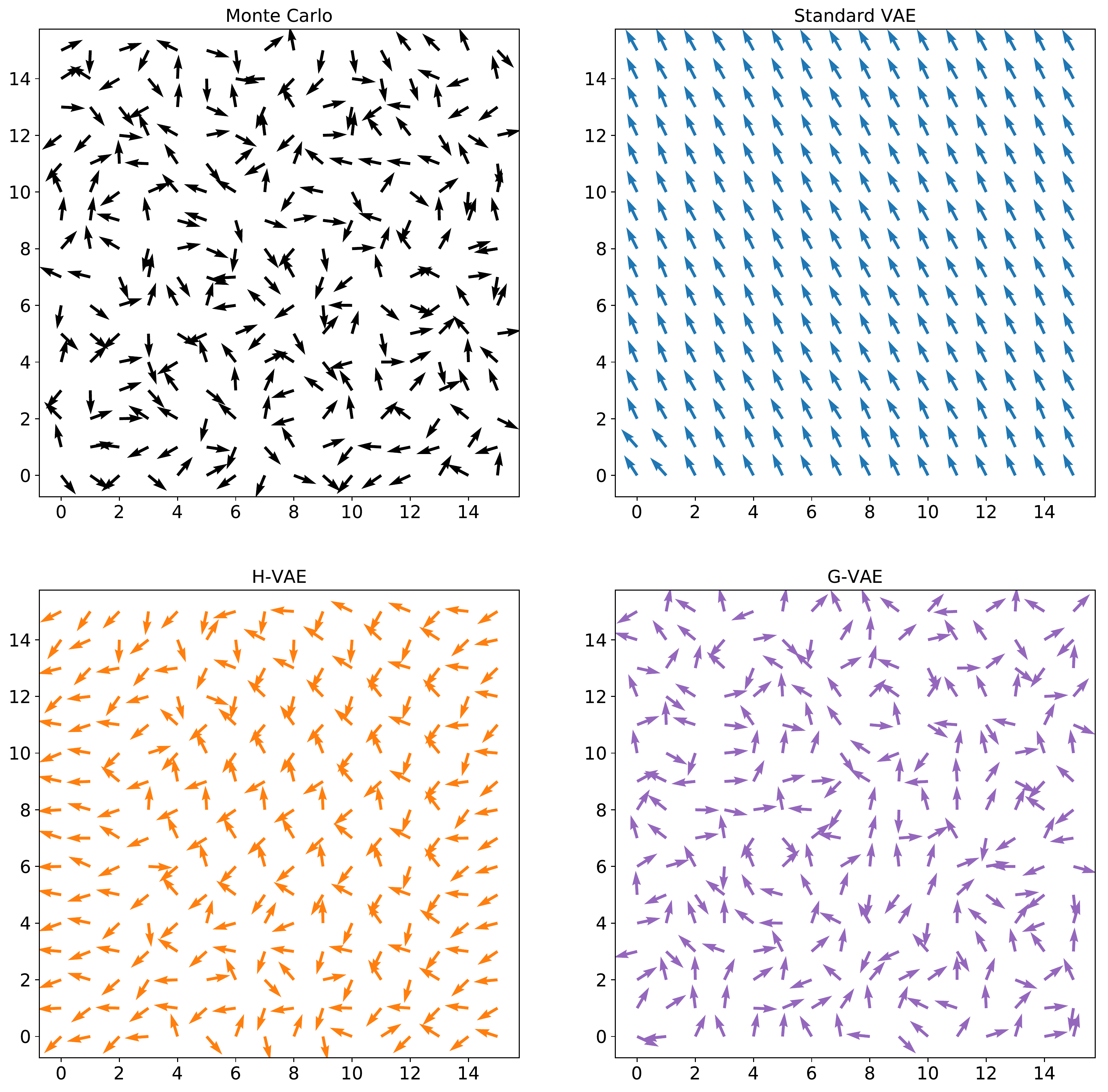}
\end{center}
\caption{View of the reconstruction with the three VAEs for a configuration at $\beta=0.1$}
\label{fig:comp_vaes_rec}
\end{figure}

The reconstruction peformance of the G-VAE is shown in Fig.~\ref{fig:comp_vaes_rec} where we show the reconstruction of a configuration at $\beta=0.1$ for the three VAEs. 
Although only qualitative, this comparison clearly shows that without the Gaussian fluctuation term, the reconstructed configurations display a much higher level of (unrealistic) order than the original Monte Carlo one.

\section{Conclusions}
\label{sec:conclusion}
We presented here an extension of VAEs for applications in the domain of computational physics. 
Even in the complex cases of phase transitions governed by modifications in the interaction among topological degrees of freedom,  we found that VAEs can detect the critical behaviour.
Results are presented showing that a physically driven modification of the VAE (H/G-VAE) improves the ability of the model in reproducing the Monte Carlo configurations. This opens the possibility of promoting VAE algorithms as instruments for the investigation of the collective degrees of freedom of Monte Carlo simulated physical configurations.

\subsubsection*{Acknowledgments}
The authors want to thank F.~Di~Renzo for the useful discussions.

\bibliographystyle{natbib}
\small
\bibliography{cristoforetti17variational}
\end{document}